# Spin Cooperated Catalytic Activities in Mn-N$_4$ based Single-atom Nanozyme: Mechanisms and a Brief Charge-spin Model

Ling Liu[1,†], Shaofang Zhang[1,†], Xinzhu Chen[1,†], Guo Li[2], Yonghui Li [2 *], Xiao-Dong Zhang [1,2 *]

1. Tianjin Key Laboratory of Brain Science and Neural Engineering, Academy of Medical Engineering and Translational Medicine, Tianjin University, Tianjin 300072, China

2. Department of Physics and Tianjin Key Laboratory of Low Dimensional Materials Physics and Preparing Technology, School of Sciences, Tianjin University, Tianjin 300350, China

3.

\*    yonghui.li@tju.edu.cn            \*    xiaodongzhang@tju.edu.cn

†: These authors have contributed equally.

# ABSTRACT

Although developing artificial enzymes has made great progress, there is still a gap between artificial enzymes and natural enzymes in catalytic performance. Designing and constructing efficient artificial biocatalysts is extremely desirable because of their high stability, low cost and easy storage. Here, we report a synthesized amino-functionalized graphene quantum dots-based manganese single atom catalyst (SAC) Mn-$N_4$, which exhibits POD-, CAT, SOD-like activities, especially the superior SOD-like activity. Recent studies have reported Mn-based SAzymes, however, the multi-enzyme mimicking catalytic mechanisms for Mn-$N_4$ are not comprehensive and in-depth enough. Therefore, we combine density functional theory (DFT) calculations and machine learning (ML) to validate the performance of the multi-enzyme mimicking activities. The DFT simulations show that Mn-$N_4$ owns a highly effective SOD in the "one-side adsorption" with a very low energy barrier of 0.077 eV, which can be attributed to variation of the preferred spin states of Mn-$O_2^{\bullet-}$ system and its "spin flip-collection lock" in the SOD-like catalytic procedure. Furthermore, spin related charge distributions on Mn-$N_4$ configurations by machine learning (ML) analysis suggest that the pattern of spin and natural charge/valence electron distribution will exhibit similarity in the structures of multiple intermediate steps of multi-enzyme mimicking activities. This work not only puts forward the catalytic mechanisms of Mn-$N_4$ SAzymes, but also provides essential guidance for future design of highly performance artificial enzymes.

# 1. Introduction

Nanozyme engineering, as an interdiscipline of chemistry, biology and physics, aim to development of artificial enzymes for precise transformations [1]. Compare with natural enzymes, many nanozymes show superior biocatalytic activities and selectivity which can be modified by molecular engineering. From the perspective of designing new catalytic materials, building a heterogeneous structure are usually a successful strategy to trigger quantum mechanical effect of materials. In a work back to 1999, surface metal atoms modified heterogeneous catalyst, Pt cations on MgO is reported [2]. Since the peroxidase-like activity of $Fe_3O_4$ nanoparticales was first reported in 2007 [3], a variety of nanozymes, including metals, metal oxides, and carbon nanomaterials have been studied as enzyme mimics due to their high stability, low cost, easy storage and multifunctionalities. After some earlier development, researchers narrow the activation sites to a single one [4] which eventually lead the emergence of M-N-C structure (a.k.a. $M-N_x$ structures) for potential nanozyme applications in electrocatalysis [5], organic transformation [6], photocatalysis [7] and biocatalysis [8].

Earlier investigations in $M-N_x$ family focus on $Fe-N_4$ and $Mn-N_4$ [9] and they show pronouncing enzyme-like behaviors. $Fe-N_x$ structure is fabricated into nanotubes and being tested for its peroxidase-mimicking activity [10]. Other reports with different metal acts as the central catalytic site are reported. Xu et al. synthesized a $Zn-N_4$ structure as antioxidant nanozyme and apply it to wound disinfection applications [11]. In 2019, $Co-N_5$, $Fe-N_5$ and $Mn-N_5$ as a new series of single-atom nanozyme are added to the $M-N_x$ family [12]. After Chen et. al. synthesized the $M-N_4$ (M=Cr, Mn, Fe, Co, Cu) materials [13], Zhang et. al. extend the $M-N_x$ family by introducing a $FeCu-N_6$ structure [14]. These new materials set up a new space for researchers to explore and adjust for new nanozymes.

Based on the structure of $M-N_x$ family of materials as nanozymes, the catalytic mechanisms are explored with theoretical tools. Nanozyme show enzyme-like reactions as oxidase, catalase, peroxidase, superoxide dismutase, glutathione peroxidase and more. Simulations focus on the binding and separation between nanozymes and superoxide/hydroxyl radical/oxygen atoms/hydrogen peroxide/etc [11,14–18]. However, the $Mn-N_4$ catalyzed reaction mechanism is still unclear. Due to the magnetic properties of Mn, its spin seems to play an important role in the reactions. So, in this work, we explore the nanozyme behaviors of $Mn-N_4$ including catalase-like, peroxidase-like, superoxide dismutase-like behaviors with spin involved critical structures. A brief data-oriented model is also provided to draw the connection between electron distribution and spin change.

On the other hand, previous studies have been devoted to exploring the relationship between spin states and catalytic activity of catalysts containing metal active sites. Lubomı́r Rulı́šek et. al. studied the detailed reaction mechanism of iron and manganese superoxide dismutase with density functional calculations on realistic active-site models and proposed that the reaction mechanisms and spin states seem to have been designed to avoid spin conversions or to facilitate them by employing nearly degenerate spin states [19]. Martin Srnec et.al. studied reaction mechanism of manganese superoxide dismutase using QM/MM approach at the density functional theory level and suggested that the oxidation of $O_2^{\bullet-}$ to $O_2$ most likely occurs by an associative mechanism following a two-state (quartet-octet) reaction profile [20]. However, the previous researches on the relationship between spin and catalysis are not deep enough, and most of the works are limited to reaction mechanism of SOD.

In this work, inspired by the synthesized of single atom catalyst $Mn-N_4$ in experiment, the structures, multi-enzyme mimicking catalytic activities of $Mn-N_4$ are explored by both experimental methods and theoretical calculations. By doping Mn single atom into the amino-functionalized graphene quantum dots, $Mn-N_4$ displays outstanding SOD-like activities. To understand the origin of the specific efficient SOD-like activities, the possible reaction mechanisms ("one-side adsorption" and "bilateral adsorption") of POD-like, CAT-like and SOD-like activities are systematically investigated by density functional theory (DFT) simulations. The simulation results reveal that $Mn-N_4$ can smoothly activate two $O_2^{\bullet-}$ to form $H_2O_2$ and $O_2$ molecules in succession in the "one-side adsorption" of SOD-like reaction, which can be attributed to variation of the preferred spin states of $Mn-O_2^{\bullet-}$ system and its "spin flip-collection lock" in the SOD-like catalytic procedure. Last but not least, spin related charge distributions on $Mn-N_4$ configurations by machine learning (ML) analysis suggest that the pattern of spin and natural charge/valence electron distribution will exhibit similarity in the structures of multiple intermediate steps of multi-enzyme mimicking activities. This work for the first time correlates the spin with multi-enzyme mimicking activities of $Mn-N_4$ and will provide essential guidance for the future design and synthesis of highly active enzyme mimics and help understand the rational catalytic mechanisms of nanozymes.

## 2. Experimental results and theoretical methods

*Preparation of Mn single-atom nanozymes*

The Mn single-atom nanozymes were synthesized by a two-step method. First, the amino-functionalized graphene quantum dots were prepared using a hydrothermal method. The procedures were as follows: 2 g of pyrene was nitrated in 160 mL of concentrated nitric acid by condensing and refluxing in an oil bath at 80 °C for 13 h. After the reaction solution was cooled to room temperature, it was washed several times by deionized water and then dried under vacuum to obtain neutral trinitropyrene. After that, 300 mg of trinitropyrene , 55 mL of deionized water and 5 mL of concentrated ammonia were added sequentially into a centrifuge tube, and the mixture was crushed in an ice water bath using a cell crusher for 2 h. The mixture was transferred to a stainless steel reactor lined with polytetrafluoroethylene and kept at 180 °C for 10 h. The larger and smaller products as well as the unreacted reagents were removed by centrifugation and dialysis to collect the final amino-functionalized graphene quantum dots. In the second step, the Mn atoms were anchored to the graphene quantum dots. The 0.01 mmol of $MnCl_2$ was dissolved in 30 mL of amino-functionalized graphene quantum dots, and the mixture was sonicated for 15 min and then freeze-dried to obtain the powder. The obtained powder was mixed with urea at a mass ratio of 1:10 and ground evenly. The mixed powder was transferred to a tube furnace and activated at 500 °C for 2 h under Ar atmosphere to obtain Mn single-atom nanozymes.

*Structural characterization of Mn single-atom nanozymes*

The atomic information and element distribution of single-atom nanomases were observed by the spherical aberration-corrected atomic resolution microscope (JEM-ARM200F, JEOL, Japan) and field-emission transmission electron microscopes (JEM-2100F, JEOL, Japan), operating at an accelerating voltage of 200 kV. X-ray diffractometer (XRD, Smartlab, Japan) equipped with Cu Kα radiation was used to detect the structure of the Mn single-atom nanozymes in the range of 10°-60°. The Al Kα-excited X-ray photoelectron spectroscopy (XPS, Axis Supra, Kratos) was employed to analyze the element composition and bonding information of Mn single-atom nanozymes. Raman spectrometer (DXR Microscope, Thermo Electron Corporation, USA) equipped with a 532 nm He-Ne laser was used to measure the structure of Mn single-atom nanozymes under a scanning wavenumber range of 800 to 3000 $cm^{-1}$.

*Enzyme-mimic activity measurements:*

The POD-like activity of Mn single-atom nanozymes was tested using the TMB substrate chromogenic kit (TELISA, SenBeiJia). According to the instructions, the TMB chromogenic solution, TMB buffer solution and TMB oxidant were were mixed into a working solution at a volume ratio of 5000:5000:4. Then 20 μL of $MnN_4$ at different concentrations (10, 20, 40, 60, 80 and 100 μg/mL) and 180 μL of working solution were mixed as the experimental group, 20 μL of the graphene carriers without Mn atom loading instead of nanozymes as the experimental control group, 20 μL of deionized water instead of nanozymes as the blank control group. The reaction mixture were added into a 96-well plate, the absorbance values at 652 nm were measured in real time using a microplate spectrophotometer (SuPerMax 3000FA, Shanghai Flash Spectrum). Finally, the catalytic rate for the first 5 min of the reaction was calculated.

Since natural catalase can catalyze the decomposition of $H_2O_2$ to $H_2O$ and $O_2$, the CAT-like activity of Mn single-atom nanozymes was assessed by using UV-visible spectrophotometer to detect the change of $H_2O_2$ content in the reaction system containing nanozymes. The 30 μL of Mn single-atom nanozymes (500 μg/mL) was mixed with 270 μL of $H_2O_2$ (40 mmol/L), and the change of absorbance in the reaction system after the addition of the nanozymes was immediately detected at 240 nm using a UV-Vis spectrophotometer. After 120 minutes of reaction, the $H_2O_2$ scavenging rate of nanozymes was determined by the equation: clearance rate = $(A_{0\ min} - A_{120\ min})/A_{0\ min}*100\%$. Finally, the decomposition rate of $H_2O_2$ by nanozymes was calculated for the first 20 min of reaction.

The SOD-like activity of the nanozymes was assayed using the Total Superoxide Dismutase Assay Kit with NBT (R22261, Yuanye). Firstly, the SOD buffer, NBT colorimetric solution and enzyme solution were mixed in the ratio of 100:30:30 to form the SOD working solution. In the assay, 160 μL of working solution, 20 μL of nanozymes at different concentrations (100, 200, 300, 400 and 500 μg/mL), 20 μL of reaction starter solution were added into the 96-well plate, respectively. The nanozymes were replaced by SOD buffer as the lighting control group, and 96-well plate was irradiated under lamp for 10 min until the lighting control group turned blue-green. The light-avoiding control group was incubated in the dark for the same time with 20 μL of SOD buffer instead of the nanozymes. Then, the absorbance at 560 nm ($A_{560}$) of each group was recorded by a microplate spectrophotometer and the absorption spectra at 300-800 nm were detected by a UV-vis spectrophotometer. The inhibition rate of nanozyme on formazan was calculated according to the formula: inhibition rate (%) = $(A_{lighting} - A_{nanozymes})/(A_{lighting} - A_{light-avoiding}) \times 100\%$. The inhibition rate of nanozyme on formazan was calculated according to the formula: The values of SOD-like activity were determined by the formula: activity (Unit) = inhibition rate/(1- inhibition rate).

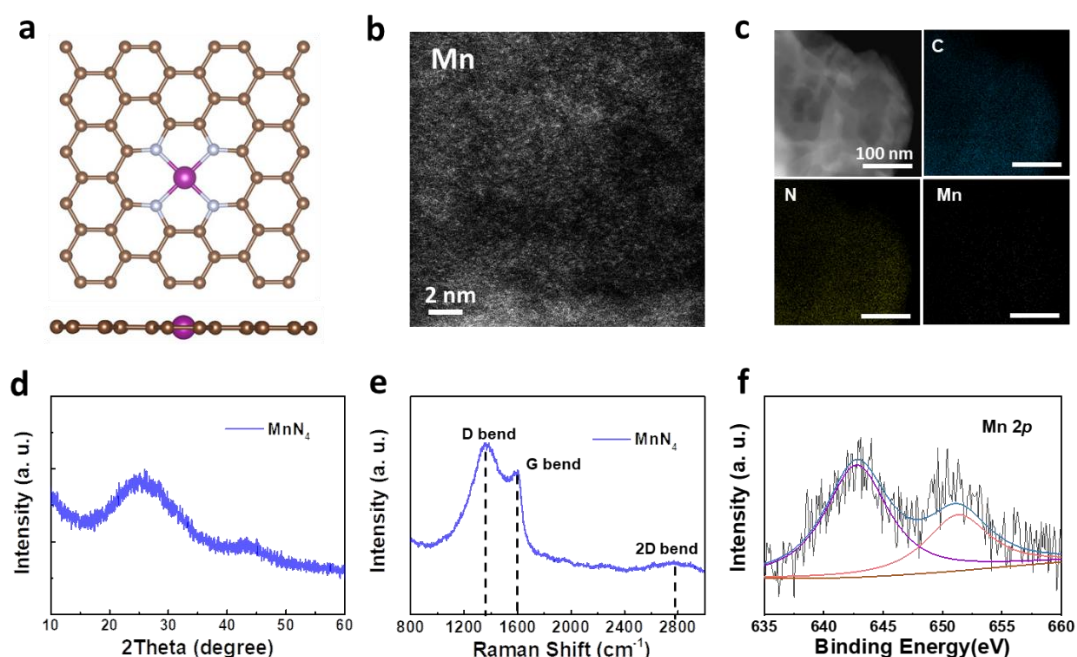

Figure 1 The structural characterization of Mn single-atom. (a) The optimized structures of MnN$_4$. (b) The AC-HAADF-STEM and (c) EDS mapping images of Mn single-atom nanozymes. The (d) XRD and (e) Raman spectra of Mn single-atom nanozymes. (f) XPS spectra of Mn single-atom nanozymes in Mn 2p region.

Since Mn-N$_4$ showed POD, CAT, and SOD-mimicking activities, mechanisms for POD-like, CAT-like, and SOD-like activities of Mn-N$_4$ catalyst are as important as the electron structure analysis of the catalyst. Therefore, two models are designed and simulated. Considering the structure of the substrate material, the first model (for catalytic energy profile simulations) is constructed as a graphene flake with its central region replaced by a single metal atom and four N atoms surrounded. The second model (for analysis of the electron structure of the catalyst) is constructed as a 12.78Å×12.30Å supercell with a Mn-N$_4$ embedded graphene sheet. To avoid the interaction between layers due to periodicity and improve the reliability and accuracy of calculations, a 15 Å vacuum distance is taken along the z-direction perpendicular to the surface of catalysts.

With the first model, intermediate states can be obtained by fully relaxing geometries of the flake with fragments/reactants attached. Intermeidate and transition states are minima and saddle points on the energy surfaces along the reaction coordinates. Based on modeled structures, all the intermediate and transition states in multi-enzyme catalytic reactions are simulated by Density Functional Theory (DFT) using the Gaussian 09 package (cite). The 6-31G(d,p) basis set is used as a balanced approach between computational workload and accuracy. DFT approximated the electron-electron interactions as a background functional of electronic density, and the functional is generally split into 2 parts, the Hartree potential and the exchange-correlation potential. The well-known B3LYP is chosen as the exchange-correlation potential. All structures are calculated at all possible spin (up to 10 in spin multiplicity) states in this work. All structures are confirmed by frequency calculations such that each intermediate state do not show any imaginary frequencies at its minimum energy and each transion state shows only one imaginary frequency. The corresponding zero-point-energy (ZPE) and thermal corrections of frequency analyses were performed at the same level of theory to obtain Gibbs free energies.

To evaluate other properties such as the stability and electronic properties of Mn-N$_4$ catalyst, the formation energies, charge density difference, electron localization function (ELF), and density of states (DOS) are investigated with the Vienna ab initio Simulation Package (VASP) [21,22]. The elemental core and valence electrons are represented using the projector augmented wave (PAW) method [23,24]. The Perdew−Burke−Ernzerhof generalized gradient approximation (GGA-PBE) functional [25] is employed to estimate the exchange−correlation potential energy. The convergence of calculations is ensured by setting the cutoff energy for the plane-wave basis at 500 eV. In addition, Brillouin zone integrations are performed according to the Monkhorst-Pack scheme[5] using a $4 \times 4 \times 1$ k-mesh. In structural relaxations, the convergence criteria for energy and force are set to $1 \times 10^{-6}$ eV and -0.01 eV/Å, respectively. The van der Waals (vdW) interaction is considered by allowing the D3 correction [26,27] and the spin polarizations are considered in all calculations. The formation energies ($E_f$) of structures are calculated by equation 1.

$$E_f = E_{total} - n_C E_C - n_N E_N - n_{Mn} E_{Mn} \quad (1)$$

Where $E_{total}$ represents the total energy of the N$_4$-G/Mn-N$_4$ configuration, $E_c$, $E_N$ and $E_{Mn}$ represent the energy of a carbon atom in graphene, half of a N$_2$ molecule energy in the gas phase, and the energy per atom for Mn, respectively. n$_C$, n$_N$ and n$_{Mn}$ are the total numbers of C, N and Mn atoms, respectively.

## 3. The structures and charge distributions of Mn-N$_4$

High-density Mn single-atom nanozyme were prepared using amino-functionalized graphene quantum dots as the substrate, and their geometric structures are displayed in Figure 1a [28]. The individual Mn atoms of the MnN$_4$ were observed by aberration-corrected high-angle annular dark-field scanning transmission electron microscopic (AC-HAADF-STEM) (Figure 1b), with bright dots representing Mn atoms marked by yellow circles. The corresponding energy-dispersive X-ray spectroscopy (EDS) mapping showed that Mn, C and N are uniformly distributed in the MnN$_4$ (Figure 1c). The X-ray diffraction (XRD) spectrum exhibited two broad characteristic peaks near 25° and 44°, which correspond to the (002) and (010) lattice planes of graphitic C, respectively, and there were no atomic aggregation states and metal oxides (Figure 1d). Similarly, the D-band peaks at 1328 cm$^{-1}$ and 2700 cm$^{-1}$ in the Raman spectrum indicated typical N-doped graphene (Figure 1e). Further, the element species and valence analysis of the MnN$_4$ were determined by X-ray photoelectron spectroscopy (XPS). The C species were divided into C=C, C-N and O-C=O bonds, and N phase consisted mainly of pyrrole N, pyridine N and graphitic N (Figure S1). The Mn 2p region was located at 642.8 and 651.3 eV, which were attributed to Mn$^{2+}$ 2p$_{3/2}$ and Mn$^{2+}$ 2p$_{1/2}$, respectively (Figure 1f).

To evaluate the structural stability of Mn-N$_4$ configuration, the electron structure level of comparisons between N$_4$-G and Mn-N$_4$ are provided including formation energies, charge redistributions, bonding properties and density of states. As calculated according to equation (1), (shown in Figure 2a), $E_f$ of the N$_4$-G structure is 1.709 eV. After the incorporation of Mn atom, the reduction of $E_f$ to -1.139 eV demonstrates that Mn-N$_4$ is energetically favored. The optimized geometries shows that all atoms in both N$_4$-G and Mn-N$_4$ structures are kept nearly in the same plane as shown in Figure 1a. Next, the charge density differences of these two models are simulated (Figure 2a and 2d). N$_4$-G in Figure 2a shows that the charges are uniformly distributed around N atoms due to the symmetry of the N$_4$ pattern. After introducing the Mn atom to the center, remarkable charge rearrangement occurs, and the Mn atom tends to lose electrons to the surrounding N atoms (Figure 2d) while carbon atoms on the graphene also donate electrons to N atoms. The electron distributions in the N$_4$-G and Mn-

$N_4$ configurations are further verified by the ELF images (Figure 2b and 2e). As shown in Figure 2b, unpaired (represented by green) and lone pair (represented by orange) electrons around every N atom extend towards the center in $N_4$-G. With the incorporation of the Mn atom, the unpaired electrons as well as the lone pair electrons of the N atoms contribute to Mn-N interactions, thus electrons around Mn behave like free electrons, which is displayed in the ELF image of Figure 2e. Besides, the electronic structures of $N_4$-G and Mn-$N_4$ are also examined by DOS analysis (Figure 2c and 2f), based on the DOS plot in Figure 2c, the electronic states are almost equally filled both in spin-up and spin-down. While the DOS pattern (Figure 2f) is markedly altered after the incorporation of the Mn atom, which eliminates the band gap of $N_4$-G and is expected to decrease the energy barrier for species' adsorption [29].

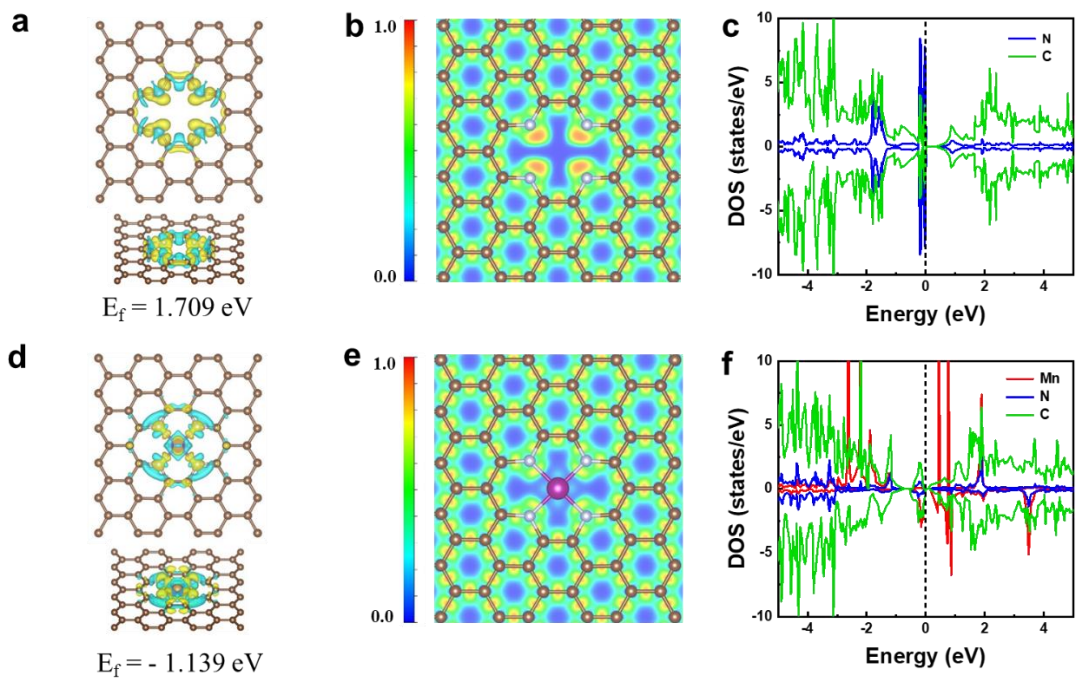

Figure 2 Top view of the charge density difference of (a) $N_4$-G and (d) Mn$N_4$ with the isosurface of 0.005 e Å$^{-3}$. cyan and yellow areas represent charge density depletion and accumulation, respectively. ELF images of (b) $N_4$-G and (e) Mn$N_4$. The density of states (DOS) of (c) $N_4$-G and (f) Mn$N_4$.

## 4. Mechanisms for enzyme-mimicking catalytic reaction

To reveal the Mn-$N_4$ multi-enzyme like activities including the POD-like, SOD-like and CAT-like performances, comprehensive DFT simulations are carried out. The structures with the lowest energy of all of the intermediate states (INT) and transition states (TS) involved in the POD-, CAT-, and SOD-like reaction activities are shown with the energy profiles in Figure 3-6 and Figure S2-5. Structures with heavy spin contamination are discarded in this work.

**4.1 Mechanisms for POD-like catalytic activities of Mn-$N_4$.**

The POD-like catalytic process is a process that decomposes $H_2O_2$ into $H_2O$. To investigate the catalytic performance of the Mn$N_4$ in enzyme-mimic catalysis, the POD- mimic reaction is first detected using colorimetric method. Compared to the graphene carriers without Mn atom loading (N/C), the Mn$N_4$ dramatically catalyzes the substrate TMB within 60 min, confirming that the excellent POD-like activity originates from the introduction of single-atom Mn (Figure 3a). The concentration-dependent catalytic rate is quantified for Mn$N_4$, which is as high as 3.31 μM/min at 10 μg/mL (Figure 3b). In addition, ESR tests reveal the significant hydroxyl radicals (•OH) scavenging ability of Mn$N_4$ (Figure 3c).

The POD-like catalytic mechanism of Mn-$N_4$ is investigated based on a variety of intermediate/transition states of Mn-$N_4$ catalyst attached to different chemical groups. The POD-like activity which can be evaluated by the 3,3',5,5'-tetramethylbenzidine (TMB) colorimetric method, is simplified in our simulation by replacing TMB by hydrogen atoms. Simulation results of different reaction pathways suggest that Mn-$N_4$ prefers the "bilateral adsorption" process of activating $H_2O_2$ by forming the Mn-O-N$_4$ structure with an active center through the heterolytic path as shown in Figure 3. The "bilateral adsorption", according to previous works [14,16], is primarily attributed to the unique "two-sided oxygen-linked" catalytic reaction path which explain the high catalytic activity of MN$_x$ due to the increase the utilization of catalysts.

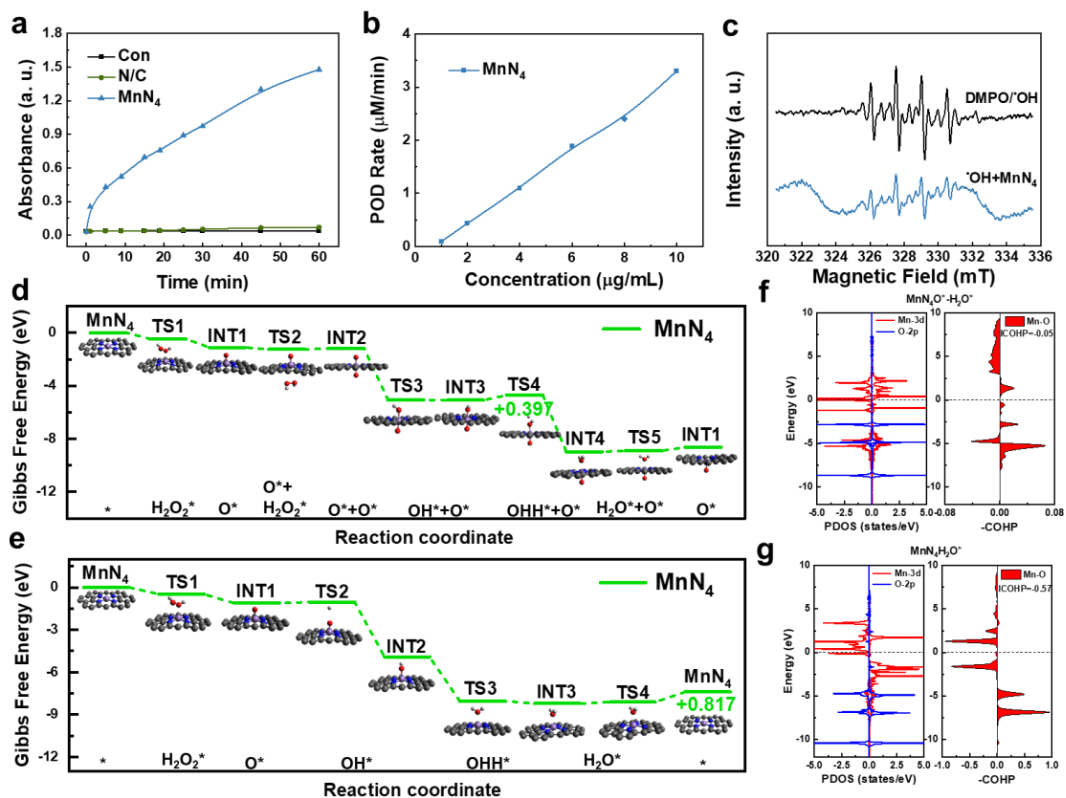

Figure 3 POD-mimic activity of Mn single-atom nanozymes. (a) Reaction-time curves of the POD-mimic reaction with TMB as substrate catalyzed by Mn single-atom nanozymes at 10 μg/mL. (b) Concentration-dependent reaction rate of Mn single-atom nanozymes catalyzing TMB within 5 minutes. (c) The •OH scavenging ability of Mn single-atom nanozymes. Calculated reaction energy profiles corresponding to the peroxidase-like activity for the heterolysis of $H_2O_2$* of MnN$_4$ corresponding to (d) bilateral mechanism and (e) one-side adsorption mechanism. The purple, blue, gray, red, and white balls represent the Mn, N, C, O, and H atoms, respectively. The DOS and COHP of (f) MnN$_4$O-$H_2O$* and (g) MnN$_4$-$H_2O$*. All 2p orbitals indicate the O 2p orbitals in $H_2O$* and all COHP is calculated for the Mn-O interaction, O indicates the oxygen atom in $H_2O$*.

The "bilateral adsorption" mechanism in the catalytic process shows a remarkably low activation energy which should be extensively discussed along the reaction pathway. As shown in Figure 3d and Figure S2a, the reaction pathway roughly goes through four steps: $H_2O_2$ molecule activation (step i), bilateral adsorptions (step ii), hydrogen attractions (step iii), $H_2O$ molecule desorption (step iv). (i) First, a $H_2O_2$ molecule can be activated by absorbing on the Mn-N$_4$ site. (ii) Then the $H_2O_2$ can be dissociated by generating a O* intermediate (INT1 in Figure 3d and S1a) and a water molecule. And another $H_2O_2$ molecule can be captured by the vacant side of the active site and forms an O*+O* intermediate state (INT2 of Figure 3d and Figure S2a) with a very small barrier of 0.072 eV. (iii) Subsequently, an O* on active site attracts two hydrogen atoms successively to form O*+OHH* intermediate state (INT4 of Figure 3d and Figure S2a), yielding one $H_2O$ molecule, which is easily surmountable for Mn-N$_4$, and this step is the rate-determining step (RDS) with a small barrier of 0.397 eV. (iv) Finally, this molecule desorbs from the catalyst with a low barrier of 0.334 eV, which is smaller than the RDS.

As shall be demonstrated, the "bilateral adsorption" mechanism bypasses the release of the last $H_2O$ molecule which corresponds to a 0.817 eV barrier which is the critical feature of reducing the activation energy along the reaction pathway. In contrast, the 0.817 eV barrier cannot be avoided in the "one-side adsorption" pathway. As shown in Figure 3e, after the formation of O* intermediate, the O* on the Mn active site can easily capture two H atoms consecutively to form OHH* (INT3 of Figure 3e). To complete the catalytic procedure, the Mn-N$_4$ must return to its initial state by releasing a $H_2O$ molecule. Therefore, the $H_2O$ desorption is the RDS with energy barrier of 0.817 eV in "one-side adsorption". If the "bilateral adsorption" behaved similarly by releasing the last H$_2$O molecule (Figure S2a), such reaction pathway had no advantage in reaction rate based preference.

Moreover, to support the "bilateral adsorption" mechanism, the crystal orbital Hamilton population (COHP) [30] analysis is applied to investigate the bonding and antibonding contribution of the adsorbed groups and bonded Mn sites. To gain more insights into origins of higher POD-like activity (lower energy barrier) in the "bilateral adsorption", the structures corresponding to the last H$_2$O molecule desorption (TS5 of Figure 3d and TS4 of Figure 3e) are studied. The COHP curves of Mn-O bond (between the Mn atom and the oxygen atom in H$_2$O) in MnN$_4$O-H$_2$O* and MnN$_4$-H$_2$O* are shown in Figure 3f and 3g, respectively. The integrated COHP (iCOHP) offers a quantitative evaluation of the bond strength in each moiety. The weakened Mn-O interaction is verified by a less negative iCOHP value in MnN$_4$O-H$_2$O* (-0.05 eV vs -0.57 eV in MnN$_4$-H$_2$O*) due to the extra O* on the other side. As shown by the DOS plot in Figure 3g, for MnN$_4$-H$_2$O*, strong p-d electronic states overlap in the low energy level (-6.9 ~ -4.9 eV) makes the bonding contributions (left panel of the COHP plot in Figure 3g) increase. While for MnN$_4$O-H$_2$O* in the DOS plot in Figure 3f, low-level electronic states (-4.7 eV) overlap by Mn 3d orbitals and O 2p orbitals makes the antibonding orbitals occupied, so more electrons are filled in the antibonding orbitals (left panel of the COHP plot in Figure 3f) far below the fermi level (-4.7 eV) for MnN$_4$O-H$_2$O* and thus the Mn-O is weakened.

Besides the •OH-producing catalytic reaction pathways are still simulated since such reaction pathway is found in similar materials based on some previous studies [11,17,18]. Namely, the homolysis of H$_2$O$_2$ is also studied with "one-side adsorption" and "bilateral adsorption" catalytic processes adopted. As shown in Figure S3a, the energy barrier of the last H$_2$O desorption (RDS) is still 0.817 eV as the POD-like process above. And for "bilateral adsorption" catalytic process in Figure S3b, the homolysis of H$_2$O$_2$* to 2OH* on the other side of the Mn-N$_4$ catalyst is the RDS with an energy barrier of 1.244 eV. As a result, the homolysis of H$_2$O$_2$ is not energetically preferred among the reaction pathways.

**4.2 Mechanisms for CAT-like catalytic activities of Mn-N$_4$.**

In a CAT-like catalytic process, H$_2$O$_2$ is decomposed to O$_2$ and H$_2$O. The catalase (CAT)-mimic activity is assessed via monitoring the process of H$_2$O$_2$ decomposition catalyzed by the MnN$_4$. As shown in Figure 4a, the MnN$_4$ presents a higher catalytic efficiency of decomposition of H$_2$O$_2$ than that of the N/C. At 120 min, the H$_2$O$_2$ clearance rate of MnN$_4$ reaches 52.4%, 17-fold higher than that of N/C (Figure 4b). Figure 4c reveals a concentration-dependent catalytic rate with 210.7 µM/min at 50 µg/mL.

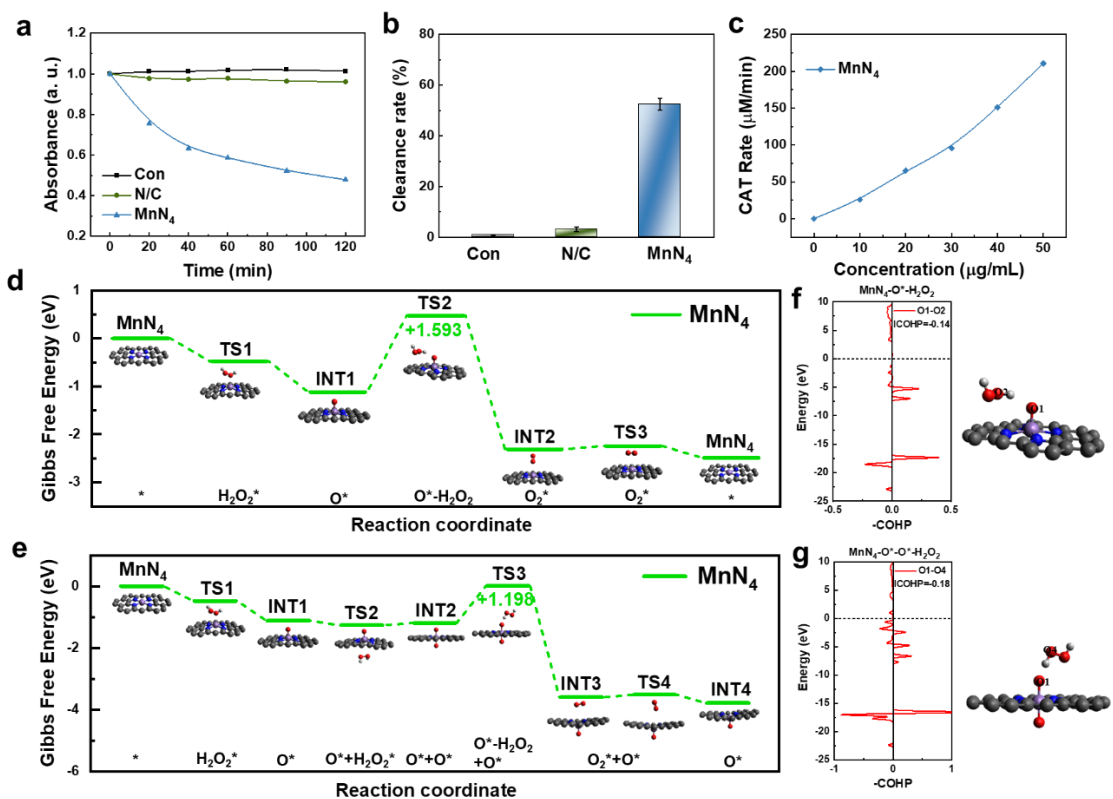

Figure 4 CAT-mimic activity of Mn single-atom nanozymes. (a) Reaction-time curves of $H_2O_2$ decomposition catalyzed by Mn single-atom nanozymes at 50 µg/mL. (b) $H_2O_2$ clearance rates of Mn single-atom nanozymes at 50 µg/mL. (c) Concentration-dependent $H_2O_2$ decomposition rate for the first 20 min. Calculated reaction energy profiles corresponding to the catalase-like activity of $MnN_4$ corresponding to (d) one-side adsorption mechanism and (e) bilateral adsorption mechanism. The purple, blue, gray, red, and white balls represent the Mn, N, C, O, and H atoms, respectively. The COHP of (f) $MnN_4O^*$-$H_2O_2$ and (g) $MnN_4O^*$-$O^*$-$H_2O_2$. All COHP is calculated for the O-O interaction, one O indicates the oxygen atom in $H_2O_2^*$, the other O indicates the oxygen atom on Mn.

The negligible CAT-like activity of Mn-$N_4$ can be attributed to the poor capacity for O* moieties on the Mn sites to activate the approaching $H_2O_2$ molecules (INT1 of Figure 4d, INT2 of Figure 4e) in both the "one-side adsorption" and "bilateral adsorption" reaction pathways. In both reaction mechanisms, the transition from $H_2O_2$ molecule adsorption to the O* intermediate state (INT1 of Figure 4d, INT2 of Figure 4e) in the CAT-like process is similar to that in POD-like process with the heterolysis of $H_2O_2^*$. So, it is important to focus on the RDS steps. In the "one-side adsorption" catalytic cycle, without extra hydrogens to react with, a high energy barrier of 1.593 eV, which is also the RDS, prevents the second $H_2O_2$ molecule from approaching to the O species on the Mn site easily. Then an oxygen atom is extracted from the $H_2O_2$ molecule to form the intermediate state $O_2^*$ (INT2 of Figure 4d) and release one $H_2O$ molecule. Although there is an $O_2$ detach from the Mn-$N_4$ surface, the energy barrier is as low as 0.066 eV.

In the "bilateral adsorption" reaction (Figure 4e and S4), the RDS has a 1.198 eV energy barrier (INT2 of Figure 5e and S4). With one O atom attached on each side, it is difficult for O* on one side to attract the subsequent $H_2O_2$ molecule. Besides, after forming $O_2^*+O^*$ intermediate (INT3 of Figure 5e and S4) and simultaneously generating a $H_2O$ molecule, $O_2$ molecule desorbs from one side of the catalyst, which is a small energy uphill step (0.084 eV). Last, the restoration of the Mn atom is unlikely to happen, since its energy barrier is as high as 1.779 eV which means that the oxygen site in O* (INT4 of Figure 4e and S4) can hardly obtain another oxygen in a $H_2O_2$ molecule to form $O_2^*$ (INT5 of Figure S4). Even when the reaction process skip the restoration, the RDS is still as high as 1.198 eV.

To understand the origin of the poor CAT-like catalytic activity (higher energy barrier) of $MnN_4$, the structures corresponding to the RDS (TS2 of Figure 4d and TS3 of Figure 4e) in "one-side adsorption" and "bilateral adsorption" are investigated. As shown in Figure 4f and 4g, the interaction between O* (TS2 in Figure 4d, TS3 in Figure 4e) on Mn site and the O atom in an approaching $H_2O_2$ molecule is weak and can be validated by small value of iCOHP. And the small iCOHP values (-0.18 and -0.14 eV) indicate the difficulty in H$_2$O$_2$ adsorption in both RDS steps. Therefore, the results of calculations suggest that the attracting for H$_2$O$_2$ molecule by the O* species on Mn site is almost unattainable either in "one-side adsorption" or "bilateral adsorption" mechanism of CAT-like catalytic activity.

**4.3 Mechanisms for SOD-like catalytic activities of Mn-N$_4$.**

In a SOD-like catalytic process, O$_2^{\bullet-}$ is decomposed to O$_2$ and H$_2$O. The quantification of superoxide dismutase (SOD)-mimic activity of the MnN$_4$ is performed based on nitro-blue tetrazolium (NBT) chromogenic principle. Figure 5a shows that the characteristic absorption of formazan disappears in the presence of MnN$_4$, indicating the superior SOD-like activity. The corresponding inhibition rate was as high as 81.2% at 50 μg/mL, much higher than 1.1% of N/C (Figure 5b). As shown in Figure 5c, the SOD-like value of MnN$_4$ is further determined to be 443 U/mg, 375 times higher than N/C.

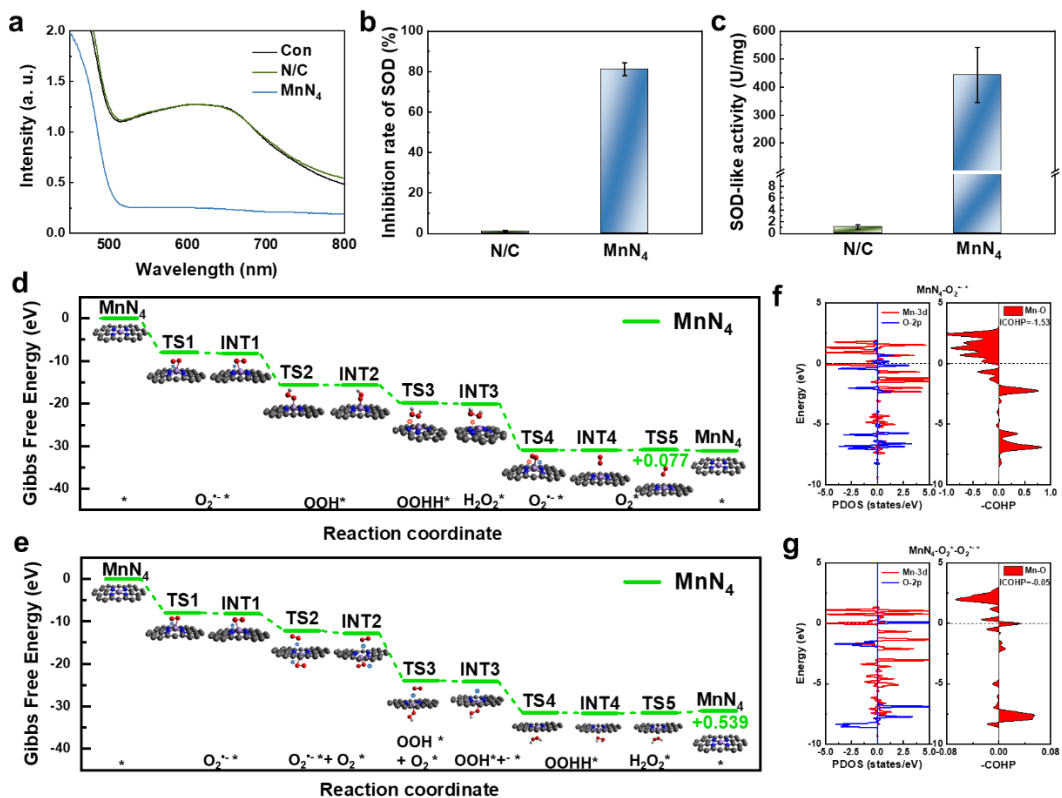

Figure 5 SOD-mimic activity of Mn single-atom nanozymes. (a) UV-vis absorption spectra of SOD-like activities with Mn single-atom nanozymes or N/C. (b) The corresponding inhibition rates of Mn single-atom nanozymes and N/C. (c) Quantification of SOD-like activity values of Mn single-atom nanozymes and N/C. Calculated reaction energy profiles corresponding to the superoxide dismutase-like activity of MnN$_4$ corresponding to (d) one-side adsorption mechanism and (e) bilateral adsorption mechanism. The purple, blue, gray, red, and white balls represent the Mn, N, C, O, and H atoms, respectively. The DOS and COHP of (f) MnN$_4$O$_2^{\bullet-}$* and (g) MnN$_4$O$_2$*-O$_2^{\bullet-}$*. All 2p orbitals indicate the O 2p orbitals of one O atom in O$_2^{\bullet-}$* and O$_2$*, respectively and all COHP is calculated for the Mn-O interaction, O indicates the oxygen atom in O$_2^{\bullet-}$* and O$_2$*, respectively.

In contrast, the most efficient SOD-like abilities are dominated by "one-side adsorption" catalytic mechanisms in Mn-N$_4$ (Figure 5d). Due to the superoxide anion radical (O$_2^{\bullet-}$) and proton (H$^+$) species around the Mn site, the valences of the transition/intermediate states vary with the charge transfers in each step. Unlike the "bilateral adsorption" mechanism (Figure 5e), the "one-side adsorption" reaction pathway roughly goes through three steps: superoxide anion radical O$_2^{\bullet-}$ adsorption (step i), consecutive proton adsorptions and H$_2$O$_2$ molecule release (step ii), O$_2$ molecule generation and desorption (step iii). (i) First, the superoxide anion radical O$_2^{\bullet-}$ readily adsorbs on the surface to form O$_2^{\bullet-}$* (INT1 in Figure 5d). The calculations suggest that adsorption of O$_2^{\bullet-}$ on the surface is highly exothermic process and is thus facile with no barriers. (ii) After two successive proton adsorptions, a H$_2$O$_2$ molecule generates (INT3 in Figure 5d) and the process is energetically and kinetically favorable in Gibbs free energy. (iii) With the desorption of the H$_2$O$_2$ molecule, another superoxide anion radical can be facilely adsorbed on the catalyst to form an O$_2^{\bullet-}$* transition state (TS4 in Figure 5d). Due to the combination of the electron from the $O_2^{\bullet-}*$ and an extra proton from the Mn-$N_4$ catalyst, an $O_2$ molecule forms (INT4 in Figure 5d) and then desorbs from the surface with a very low energy barrier of 0.077 eV, which is the RDS in the "one-side adsorption" mechanism. Finally, the Mn-$N_4$ recovers to the its initial status.

On the other hand, the RDS in the "bilateral adsorption" catalytic mechanism is the restoration of Mn-$N_4$ (TS5→Mn-$N_4$) with a 0.539 eV energy barrier. Though, in the proposed "bilateral adsorption" mechanism, an oxygen molecule is released before the release of an $H_2O_2$ molecule, the energy increase in the restoration step cannot be avoided. As shown in Figure 5f and 5g, the Mn-O interaction in the first adsorption of $O_2^{\bullet-}$ is weakened by the adsorption of the second $O_2^{\bullet-}$. Such weakening can be represented by the iCOHP of Mn-O which is reduced from -1.53 eV (in Mn-$N_4$-$O_2^{\bullet-}*$) to -0.05 eV (in $O_2*$-$^-$Mn-$N_4$-$O_2^{\bullet-}*$). In addition, it can be seen from the DOS (Figure 5f and 5g) that d-p hybridization contributes less to bonding orbitals at deep levels for $O_2*$-$^-$Mn-$N_4$-$O_2^{\bullet-}*$. Therefore, the decreased Mn-O interaction is responsible for the converting from adsorbing $O_2^{\bullet-}*$ to desorbing $O_2*$ on one-side of the Mn-$N_4$ surface due to the extra $O_2^{\bullet-}*$ on the other side.

**4.4 "Spin flip-collection lock" in the SOD-like catalytic procedure**

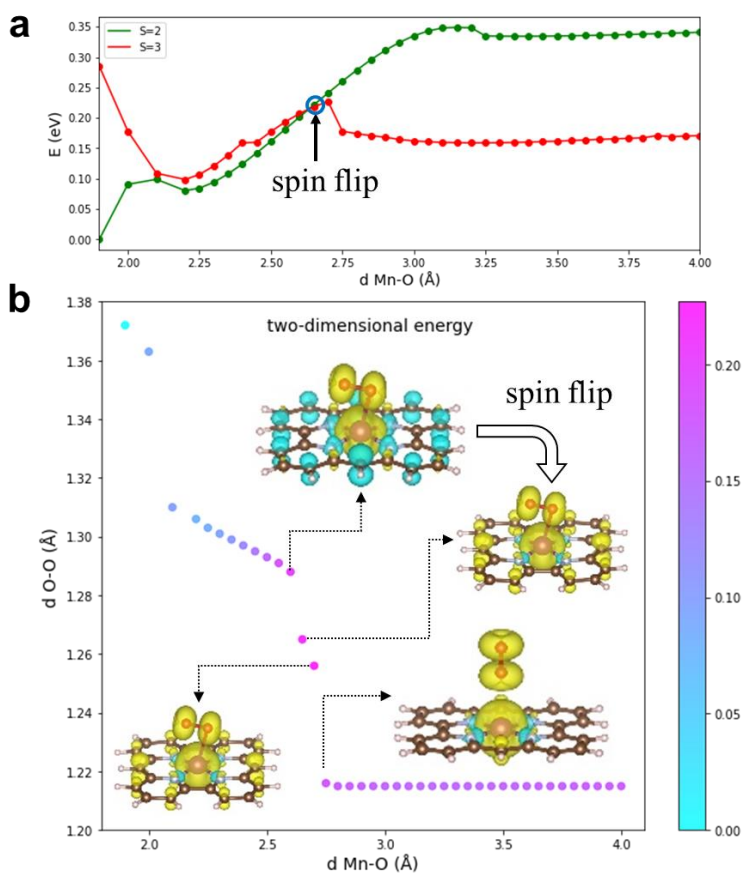

Figure 6 The energy profiles along with the Mn-O distances corresponding to the converting from superoxide anion radical $O_2^{\bullet-}$ to $O_2$ molecule on the active site of MnN$_4$ at two spin states (S=2, 3). Spin density of MnN$_4$-$O_2^{\bullet-}*$ and MnN$_4$-$O_2*$ are also plotted (yellow stands for spin up and cyan for spin down).

In a more detailed electron structure analysis, the mystery behind the adsorption of $O_2^{\bullet-}$ is close related to states of spin which are dynamically related to the small changes in the Mn-O distance. The attached $O_2^{\bullet-}$ cannot be released immediately in the form of an oxygen molecule. Such forbidden can be demonstrated as a "spin flip-collection lock" mechanism.

To demonstrate the critical moment of the "spin flip-collection lock", a series of relaxed scans are performed based on the constrained optimization . In each step of the process of the scan, the Mn-O distance is fixed at a particular value and the other variables are relaxed. Initially, one oxygen atom in the superoxide anion $O_2^{\bullet-}$ is at a distance of 1.9 Å from the manganese, the Mn-O distance increases by 0.05 Å with each scanning step, eventually moving to a distance of 4 Å from each other. As shown in Figure 6 and Figure S5, the simulated energy profiles along the process from adsorbing $O_2^{\bullet-}*$ to desorbing $O_2*$ defined by the Mn-O distance at different spin states are depicted.

As shown in Figure 6a, at a Mn-O distance of ~ 2.65 Å, "spin flip lock" unlocks, namely, from this crossing point on (Figure 6a), the spin state of the more energetically favorable structures then changes from quintet (S=2) to septet (S=3) state. The quintet- septet crossing is the energy uphill point on the profile with an activation barrier of about 0.218 eV with respect to the initial quintet state at 1.9 Å, suggesting that this "spin flip" needs to overcome an energy barrier. Moreover, as shown in Figure 6b, at a Mn-O distance of ~ 2.60 Å, the spin density shows that Mn and $O_2^{\bullet-}$ on the site own the same spin direction, while the four coordinated N atoms and some neighboring C atoms on the catalyst surface display the opposite spin of Mn and $O_2^{\bullet-}$. When the "spin-flip" occurs at a Mn-O distance of ~ 2.65 Å, the spin direction of these C atoms on the catalyst flips to the same as Mn and $O_2^{\bullet-}$ moiety. When the Mn-O distance is elongated to 2.70 Å, the spin direction of all the atoms on Mn-$N_4$ catalyst almost keeps unchanged, and as the Mn-O distance continues to elongate to 2.75 Å, an $O_2$ molecule generates and collects the spin densities from catalyst. In addition, the "spin flip" also brings changes in O-O bond length in $O_2^{\bullet-}$ moiety. In Figure 6b, the O-O bond lengths of $O_2^{\bullet-}*$ moiety become shorter (from 1.372 to 1.288 Å) as the Mn-O distance increases from 1.90 to 2.60 Å, and the "spin flip" keeps locking. Then, the O-O distance of $O_2^{\bullet-}*$ moiety decreases to 1.265 Å when the "spin-flip" occurs. Subsequently, the energy rises to the highest at a Mn-O distance of ~ 2.70 Å, and the O-O bond length of $O_2^{\bullet-}*$ moiety decreases to 1.256 Å. After that, the O-O bond length dramatically reduces to 1.216 Å with the Mn-O distance increases to 2.75 Å, an $O_2$ molecule generates simultaneously and the energy dramatically drops after the highest point (0.227 eV) and then tends to flatten (Figure 6a). Therefore, the simulations indicate that the conversion from adsorbing $O_2^{\bullet-}*$ to $O_2$ desorption needs to overcome a small energy barrier of 0.227 eV due to the presence of this "lock", as the step i of SOD-like reaction pathway described above.

**4.4 Spin related charge distributions on Mn-$N_4$ configurations by machine learning (ML) analysis.**

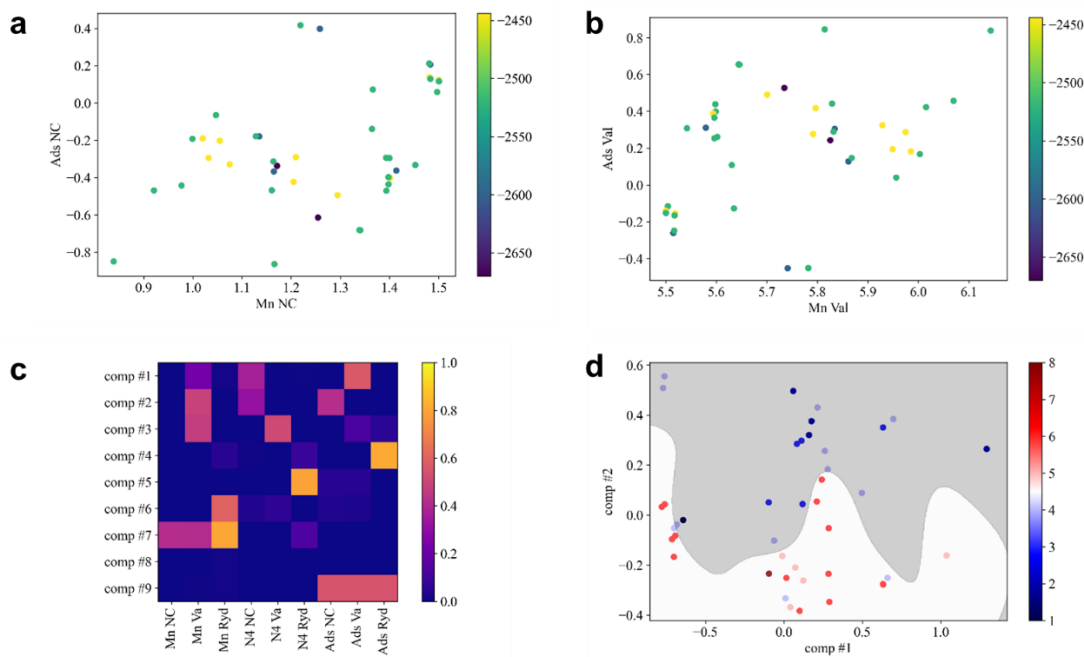

Figure 7 Collected data of energy as a function of (a) natural charge and (b) valence of Mn and adsorbates. (c) Normalized coefficient in PCA converting row-wise represented. (d) Classification of zones of low spin vs high spin based on PCA reduced first and second components.

In fact, such spin related energy shifts are broadly seen in many structures. Based on the calculation results of Mn single atom catalyst, we try to analyze from the perspective of natural population analysis (NPA), and explore the possible patterns of charge distribution, energy and spin. Based on all the calculated NPA results, the total Natural Charge (NC), Valence (Val) and Rydberg (Ryd) of Mn atom, $N_4$ structure and adsorbed groups have been selected, for a total of 9 sets of data. The number of valance electrons of the adsorbed groups is difficult to be compared horizontally because of the different groups, so the total number of valence electrons at 0 valence of each atom in the group is subtracted from this term. According to the generality of NPA, NC, Val and Ryd reflects the number of equivalent net charges of an atom in the system, the number of equivalent valence electrons an atom owns in the system and equivalent extra-valence shell of an atom because of the influence of other atoms in the system, respectively.

With the help of NPA dataset (see the Table below), one can sketch the energy distribution over various structures as a function of Mn NC and Ads NC (Figure 7a). Structures at different energies may have similar Mn NC and Ads NC, indicating that their electronic structures can show similar distribution despite different adsorbates. In addition, a similar conclusion can be given for valence electrons (Figure 7b). This phenomenon is positive evidence for the catalyst, because the unique electronic structure of the catalyst needs to be reflected in the adsorption, desorption and multiple intermediate steps/structures. Then there will be similar natural charge or valence electron distribution appearing in the structures of multiple intermediate steps.

Finally, based on NPA data, principle component analysis (PCA) were employed. At the data level, we did not perform normalization because the NPA results can be understood in terms of the number of charges, and normalization would destroy the charge number interpretation. The results showed that PCA could compress nine variables into two variables and still maintain an explained variance ratio of 97 percent. The principle components comp #1 and comp #2 transformed by PCA are shown in Figure 7c. comp #1 is mainly composed of Mn Val, Ads Val and $N_4$ NC, among which Ads Val is the most important. While comp #2 is mainly composed of Mn Val, Ads NC and $N_4$ NC, among which Mn Val is the most important. It can be seen that the first two variables of PCA describe the NPA of the electronic structure from the adsorbates and the metal Mn respectively, and the NC reflected by the $N_4$ structure is also very important, which needs to be described to a certain extent in order to comprehensively understand the change of the electronic structure in the catalytic processes. In addition, a support vector machine (SVM) classification model (Figure 7d) for spin described by comp #1 and comp #2 was established according to the composite indexes comp #1 and comp #2 transformed by PCA as descriptors. Rbf kernel (gamma=15) was adopted in the classifier and the regularization parameter is set to 10. And the regularization parameter is set to 10. If we call the spin below 4.5 a low spin state and above 4.5 a high spin state, depending on comp #1 and comp #2, the low and high spin states can be roughly separated by NPA results. The high spin states are mainly concentrated in the region where the sum of Mn Val, Ads NC and $N_4$ NC is negative. There are three expansions to high value comp #2 with different comp #1, and the whole classification boundary is w-shaped. This SVM classification roughly reflects the pattern of spin and charge/valence electron distribution.

## 5. Conclusions

In summary, we present a SAzyme with multi-enzyme mimicking activities by using amino-functionalized graphene quantum dots-derived Mn-$N_4$ single atom catalyst. Owing to the variation of spin states for Mn, the excellent SOD-like performance is demonstrated by experimental and theoretical results. We then systematically investigate catalytic mechanisms for POD-, CAT-, SOD-like activities via density functional theory (DFT) simulations, and it is disclosed that the excellent SOD-like activities can be attributed to its "spin flip-collection lock" in the "one-side adsorption" SOD-like catalytic process. Finally, spin related charge distributions on Mn-$N_4$ configurations by machine learning (ML) analysis suggest that the pattern of spin and natural charge/valence electron distribution will exhibit similarity in the structures of multiple intermediate steps of multi-enzyme mimicking activities.

This work links the spin with multi-enzyme mimicking activities of Mn-$N_4$ for the first time. We believe it will provide essential guidance for the future design and synthesis of highly active enzyme mimics and help understand the underlying catalytic mechanisms of nanozymes.

## Supplemental Information

In addition, the COHP curves of Mn-O bond in Figure S3c and S3d also elucidate the high barrier for the homolysis of $H_2O_2*$ to $2OH*$ on the other side. It can be seen that the iCOHP is decreased from from -0.63 eV (Mn-$N_4$-$H_2O_2*$) to -0.13 eV (Mn-$N_4$-OH*-$H_2O_2*$), so the weakened Mn-O interaction is also verified by a less negative iCOHP value of Mn-$N_4$-OH-$H_2O_2*$ as well as the anti-bonding contributions of Mn-d and O-p orbitals overlap in the low-level electronic states (-6.2 eV). Therefore, the weakened Mn-O (oxygen atom in $H_2O_2$ molecule) interaction is account for the poor capacity of activating the second $H_2O_2$ molecule for Mn site on the other side. In addition, another probability of homolysis of $H_2O_2*$ catalytic path has also been proposed (Figure Sxa), the RDS is still the last desorption of $H_2O$ molecule. Therefore, all the DFT simulated possible POD-like pathways suggest that the Mn-$N_4$ preferentially forms the Mn-O* structure for the heterolysis of $H_2O_2*$ in a way of "bilateral adsorption" because of the more kinetically favorable reaction path.

On the other hand, the RDS in the "bilateral adsorption" catalytic mechanism is also investigated. For simplicity, we denote the four steps as follows: superoxide anion radical $O_2^{•-}$ adsorption (step i), bilateral adsorptions (step ii), $O_2$ molecule release (step iii), $H_2O_2$ molecule generation and desorption (step iv). (i) The initialized state has an attached superoxide anion radical $O_2^{•-}$ on the surface to form $O_2^{•-}*$ (INT1 in Figure 5e) with no barriers, which is similar to that in "one-side adsorption". (ii) Then, another superoxide anion radical $O_2^{•-}$ can be adsorbed on the vacant side of the active site to form an $O_2^{•-}* + O_2^{•-}*$ intermediate state (INT2 in Figure 5e) with a favorable energy downhill of about -5 ~ -4 eV. (iii) After "bilateral adsorption" process, an $O_2$ molecule has been formed and the Mn-$N_4$ catalyst has stored extra electrons. As shown in Figure 5f and 5g, the iCOHP of Mn-O (oxygen atom in $O_2^{•-}*$ and $O_2*$, respectively) interaction is weakened from -1.53 eV (Mn-$N_4$-$O_2^{•-}*$) to -0.05 eV ($O_2*$-•Mn-$N_4$-$O_2^{•-}*$). In addition, it can be seen from the DOS (Figure 5f and 5g) that d-p hybridization contributes less to bonding orbitals at deep levels for $O_2*$-•Mn-$N_4$-$O_2^{•-}*$. Therefore, the decreased Mn-O interaction is responsible for the converting from adsorbing $O_2^{•-}*$ to desorbing $O_2*$ on oneside of the Mn-

$N_4$ surface due to the extra $O_2^{\cdot-}*$ on the otherside. (iv) Subsequently, INT2 can react with a proton in the water to form $O_2*-\,^{\cdot}Mn\text{-}N_4\text{-}OOH*$ where one electron lost and thus leads to a better desorption capacity of $O_2$ molecule on the other side. The generated $^{-\cdot}*\text{-}Mn\text{-}N_4\text{-}OOH*$ intermediate state (INT3 in Figure 5e) is still able to attract the second proton to generate $H_2O_2$ molecule (INT4 in Figure 5e). Finally, the simulations show that the desorption of $H_2O_2$ is the RDS with a moderate energy barrier of 0.539 eV, which is higher than that in "one-side adsorption". Therefore, the simulation results suggest that the SOD-like process of Mn-$N_4$ exhibits an outstanding performance ($E_b$ = 0.077 eV) in the "one-side adsorption" reaction pathway, even though the energy barrier is not high in "bilateral adsorption".

For the spin state S=0, the energy is monotonically increasing, showing that $O_2^{\cdot-}$ is sticking to the Mn-$N_4$ surface all the time. And for the spin state S=1, all the energetically favorable structures are found to have heavy spin contamination. Therefore, the rest two spin states (S=2 and 3) are taken into consideration.

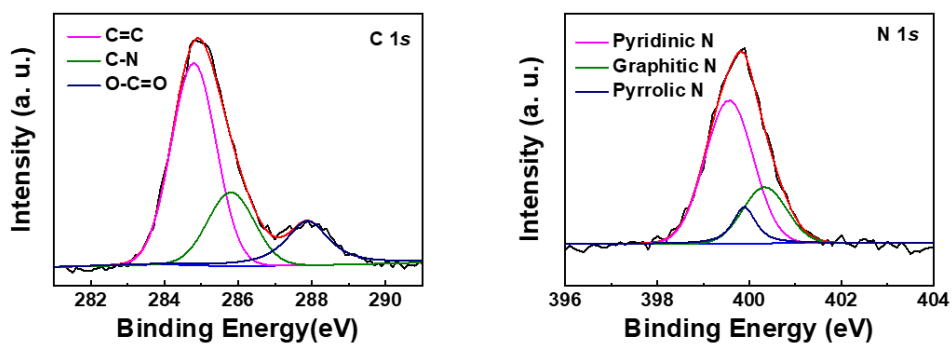

Figure S1 XPS spectra of Mn single-atom nanozymes in (a) C 1s and (b) N 1s regions.

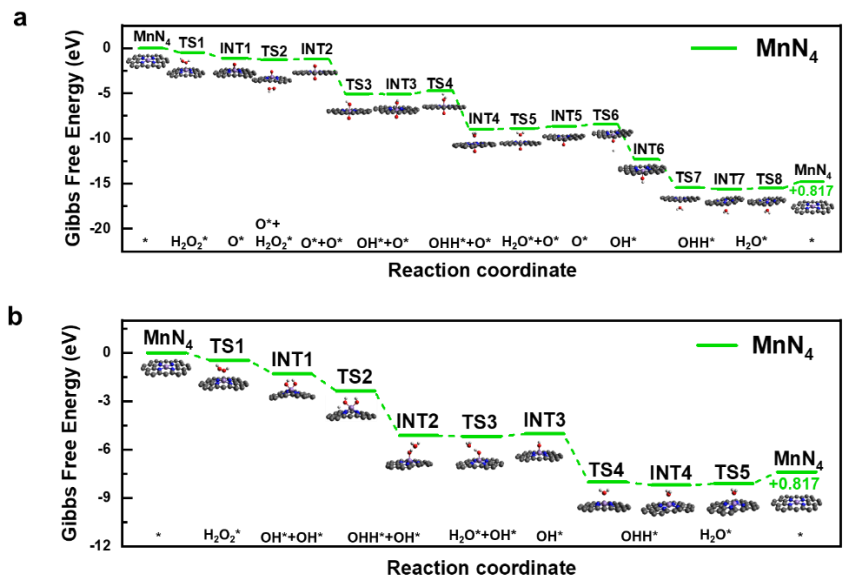

Figure S2 Calculated other possible reaction energy profiles corresponding to the peroxidase-like activity for (a) the heterolysis of $H_2O_2*$ of $MnN_4$ corresponding to bilateral adsorption mechanism. (b) the homolysis of $H_2O_2*$ of $MnN_4$ corresponding to one-side adsorption mechanism and The purple, blue, gray, red, and white balls represent the Mn, N, C, O, and H atoms, respectively.

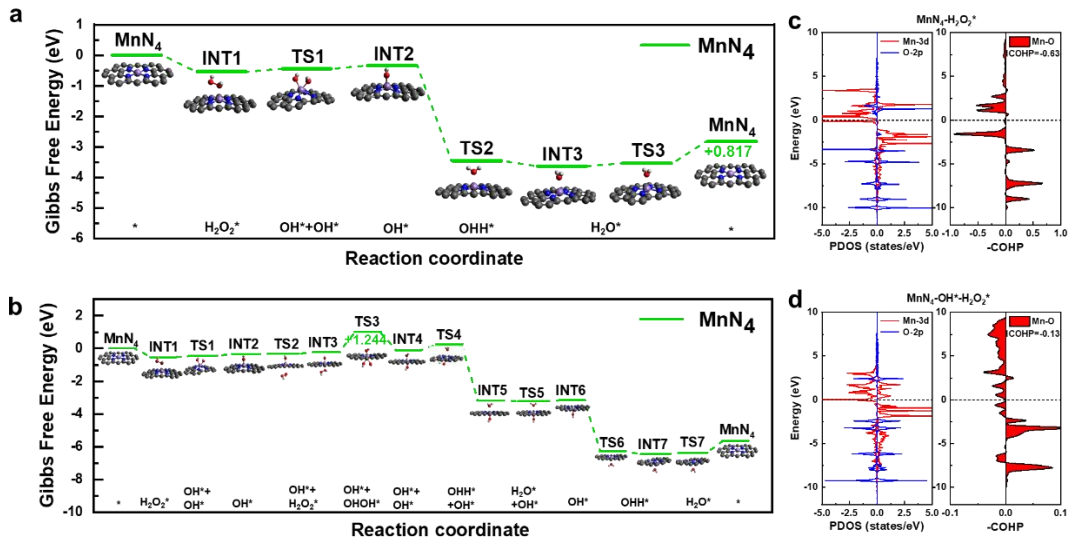

Figure S3 Calculated reaction energy profiles corresponding to the peroxidase-like activity for the homolysis of $H_2O_2^*$ of $MnN_4$ corresponding to (a) one-side adsorption mechanism and (b) bilateral adsorption mechanism. The purple, blue, gray, red, and white balls represent the Mn, N, C, O, and H atoms, respectively. The DOS and COHP of (c) $MnN_4$-$H_2O_2^*$ and (d) $MnN_4$-OH-$H_2O_2^*$. All 2p orbitals indicate the 2p orbitals of one O atom in $H_2O_2^*$ and all COHP is calculated for the Mn-O interaction, O indicates one of the oxygen atoms in $H_2O_2^*$.

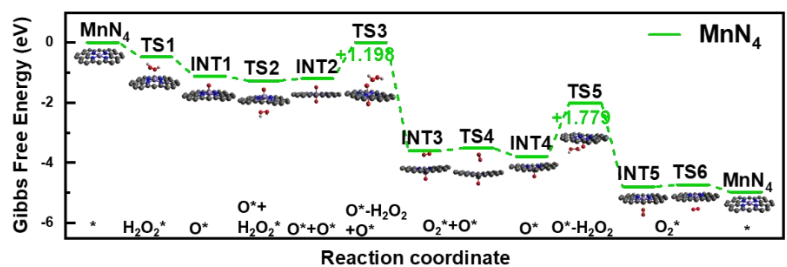

Figure S4 Calculated other possible reaction energy profiles corresponding to the catalase-like of $MnN_4$ corresponding to bilateral adsorption mechanism. The purple, blue, gray, red, and white balls represent the Mn, N, C, O, and H atoms, respectively.

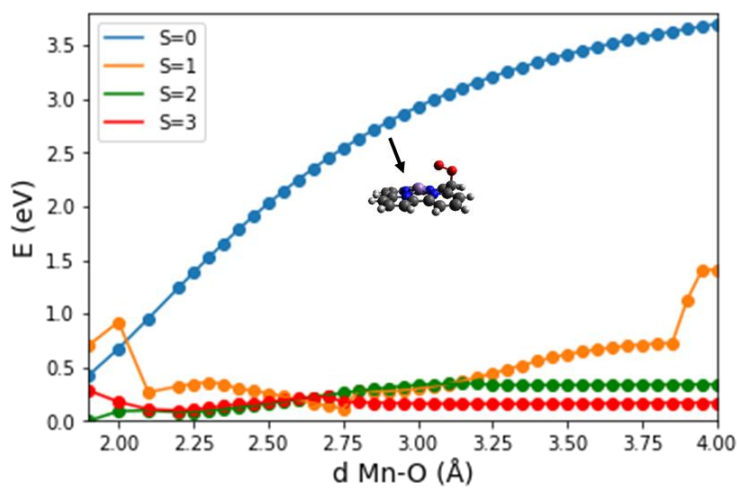

Figure S5 The energy profiles along with the Mn-O distances corresponding to the converting from superoxide anion radical $O_2^{\cdot-}$ to $O_2$ molecule on the active site of $MnN_4$ at different spins.
19